# Co-existence of Whistler Waves with Kinetic Alfven Wave Turbulence for the High-beta Solar Wind Plasma

Manish Mithaiwala, Leonid Rudakov[1], Chris Crabtree, and Gurudas Ganguli
Plasma Physics Division, Naval Research Laboratory, Washington, DC 20375-5346
[1]Icarus Research Inc., P.O. Box 30780, Bethesda, MD 20824-0780

It is shown that the dispersion relation for whistler waves is identical for a high or low beta plasma. Furthermore in the high-beta solar wind plasma whistler waves meet the Landau resonance with electrons for velocities less than the thermal speed, and consequently the electric force is small compared to the mirror force. As whistlers propagate through the inhomogeneous solar wind, the perpendicular wave number increases through refraction, increasing the Landau damping rate. However, the whistlers can survive because the background kinetic Alfven wave turbulence creates a plateau by quasilinear diffusion in the solar wind electron distribution at small velocities. It is found that for whistler energy density of only $\sim 10^{-3}$ that of the kinetic Alfven waves, the quasilinear diffusion rate due to whistlers is comparable to KAW. Thus very small amplitude whistler turbulence can have a significant consequence on the evolution of the solar wind electron distribution function.

## 1) Introduction

Measurements of the short wavelength range of solar wind turbulence spectrum, below the proton gyro-radius, $k_\perp \rho_i \geq 1$ down to the electron gryoradius $k_\parallel \rho_e \sim 1$, have revealed the existence of low frequency kinetic Alfven wave (KAW) fluctuations [Alexandrova et al., 2009, Sahraoui et al., 2009]. The gyroradius of electrons (ions) is denoted by $\rho_{e(i)}$, and the perpendicular (parallel) wavenumber with respect to the magnetic field by $k_{\perp(z)}$. It was thought, based on theoretical arguments, that KAW should be damped well before $k_\parallel \rho_e \sim 1$ [Gary and Smith, 2009; Podesta et al., 2010]. Thus it was argued that whistler waves are more likely to be the short wavelength fluctuations of the turbulence spectrum [Stawicki et al., 2001]. This ambiguity arises because of difficulty in resolving the wave frequency in the observational data. However, theoretically the measurement of KAW does not preclude the existence of whistler turbulence. Indeed, recent measurements of differences between the observed scaling of the second order structure functions of the magnetic field indicate the presence of other modes [Chen et al., 2010].

In the collisionless solar wind there is no reason that the distribution functions should be Maxwellian. Non-Maxwellian features, such as high-energy tails, are often observed in the solar wind plasma [Nieves-Chinchilla and Viñas, 2008]. Using the measured turbulence spectrum of KAW Rudakov et al. [2012] showed that quasilinear diffusion should lead to a flattened (plateau-like) distribution for the solar wind electrons at velocities less than the thermal velocity $v/v_{te} < 1$, which reduces the Landau damping of KAW, while the tail ions develop a step-like distribution. It was further shown that such an ion distribution is unstable and can lead to the generation of whistler waves. In this way some part of kinetic Alfven wave flux is converted into whistlers [Rudakov et al., 2012].

In this article it will be shown that without the formation of the plateau, whistler waves would also be strongly Landau damped. Both whistlers and KAW have the Landau resonance in the same portion of the velocity distribution function. This is due to the similarity of the dispersion relation and field equations of whistlers and KAW in a high beta plasma. Also whistlers may scatter by KAW, similar to the three-wave interaction of KAW [Rudakov et al., 2011]. The three-wave interaction provides a natural mechanism for evolution of whistler turbulence.

The damping rate of whistlers increases with perpendicular wavenumber $k_\perp$, and $k_\perp$ increases by refraction as whistlers propagate through the inhomogeneous solar wind. It would then be expected that whistlers would damp when sufficiently large $k_\perp$ has developed, independently of any cascade process that may exist for whistler turbulence [see e.g. Shaik, 2009 and references therein]. However, as it will be shown in Section 3, development of a plateau-like electron distribution under KAW as well as whistler turbulence leads to diminished Landau damping rate of whistlers. Though the amplitude of whistler turbulence is unknown, by comparing the quasilinear diffusion coefficient from KAW to whistler turbulence, an estimate can be made.

## 2) Theoretical Background

The general dispersion relation for kinetic Alfven waves is determined from the electron current $\vec{j}_k$. Coordinates are chosen such that $k_y \equiv 0$ throughout. The components of the electric field are found from Ampere's law, $\vec{j}_k = ck^2\vec{A}_k/4\pi$, in the Coulomb gauge $k_x A_{kx} + k_z A_{kz} = 0$. Throughout this article, all wavenumbers are normalized so that $\bar{k} \equiv kc/\omega_{pe}$. The cyclotron frequency is $\Omega_{e(i)} \equiv \mp eB_0/m(M)c$, the plasma frequency $\omega_{pe(i)} \equiv 4\pi e^2 n_0/m(M)$, electron plasma beta $\beta_{\perp,\parallel} = 2\omega_{pe}^2 v_{tze}^2/\Omega_e^2 c^2$, $v_{tez,\perp}^2 \equiv T_{ez,\perp}/m$ where $T_e$ is the electron temperature, $n_0$ is the electron density, and $m$ is the electron mass. The proton plasma beta is $\beta_{\perp,\parallel i} \equiv 8\pi n_0 T_{\perp,\parallel i}/B_0^2$ thermal velocity $v_{ti} \equiv \sqrt{T_i/M}$, where $T_i$ is the proton temperature, $\rho_i \equiv v_{ti}/\Omega_i$ is the ion gyroradius, and $M$ is the proton mass. With these definitions the components of the susceptibility tensor $\chi$ can directly be written to lowest order in $\lambda \equiv \bar{k}_\perp^2 \beta_\perp/2$ for $\lambda \ll 1$ [Stix, 1992], which includes $E \times B$, polarization, diamagnetic, and centrifugal drift currents. From the component's of Ampere's law, for an isotropic temperature $\left(T_\parallel \equiv T_\perp\right)$, and under the restriction of long wavelength waves $\left(\bar{k}^2 \ll 1\right)$,

$$\bar{k}^2\left(E_x + ik_x\varphi\right) = -i\frac{\omega}{\Omega_e}E_y + \frac{\omega^2}{\Omega_e^2}\frac{2}{\bar{k}_\perp^2\beta_{\perp i}}E_x, \tag{1}$$

$$\bar{k}^2 E_y = \frac{i\omega}{\Omega_e}E_x - \bar{k}_\perp^2\beta_\perp\left(1 - \frac{\beta_\perp}{\beta_\parallel}\frac{\overline{Z}'}{2}\right)E_y + \frac{i\omega}{\Omega_e}\frac{\bar{k}_\perp}{\bar{k}_z}\frac{\beta_\perp}{\beta_\parallel}\frac{\overline{Z}'}{2}E_z, \tag{2}$$

$$\overline{k}^2\left(E_z + ik_z\phi\right) = -\frac{i\omega}{\Omega_e}\frac{\overline{k}_\perp}{\overline{k}_z}\frac{\beta_\perp}{\beta_\parallel}\frac{\overline{Z}'}{2}E_y - \frac{2\omega^2}{\Omega_e^2\overline{k}_z^2}\frac{\beta_\perp}{\beta_\parallel}\frac{\overline{Z}'}{2}E_z. \tag{3}$$

The plasma dispersion function is defined as, $\overline{Z}(\zeta_e) = \frac{1}{\sqrt{\pi}}\int dv\frac{\partial f_{0e}/\partial v}{v - \zeta_e}$, for an arbitrary distribution function, with argument $\zeta_e = \omega/\sqrt{2}\,k_z v_{tze}$. Here and below $f_{0e}$ normalized as

$\int f_{0e}(v_z)dv_z = 1$. When $f_{0e}$ is a Maxwellian, $\overline{Z}$ is equivalent to the well know Z-function [Fried

and Conte, 1961].

The dispersion relation, and field equations for KAW is found beginning from the x-component Ampere's Law (1), substituting the gauge condition for the field $E_x = -\left(E_z + ik_z\phi\right)\overline{k}_z/\overline{k}_\perp - ik_x\phi$, and subtracting z-component Ampere's Law (3) gives an auxiliary equation for the fields

$$i\frac{\omega}{\Omega_e}\frac{\overline{k}_\perp}{\overline{k}_z}\left(1 + \frac{\beta_\perp}{\beta_\parallel}\frac{\overline{Z}'}{2}\right)E_y - \frac{2\omega^2}{\Omega_e^2}\frac{1}{\overline{k}_z^2\beta_{\perp i}}\frac{\overline{k}_\perp}{\overline{k}_z}E_x = -\frac{2\omega^2}{\overline{k}_z^2\Omega_e^2\beta_\parallel}\frac{\overline{Z}'}{2}E_z. \tag{4}$$

Substituting $E_z$ from (4) in the y-component (2) when $\overline{k}_z^2 \ll \overline{k}^2$,

$$\left(1 - \beta_\perp\frac{\overline{Z}'}{2}\right)\overline{k}_\perp^2 E_y = i\frac{\omega}{\Omega_e}E_x\left(1 + \frac{\beta_\perp}{\beta_{\perp i}}\right), \tag{5}$$

and the z-component (3),

$$\left(\frac{2\omega^2}{\Omega_e^2}\frac{1}{\overline{k}_\perp^2\beta_{\perp i}} - \overline{k}_z^2\left(\frac{\beta_\parallel}{\beta_{\perp i}} + 1\right)\right)E_x = i\frac{\omega}{\Omega_e}E_y. \tag{6}$$

For short perpendicular wavelengths of KAW, $E_x \simeq -ik_x\phi$ is electrostatic and $E_z = ik_z\phi\beta_\perp/\beta_{\perp i}$.

Thus $|E_x|^2 \gg |E_z|^2$, and from (5) $|E_y|^2 \ll |E_x|^2$, $|E_y|^2 \sim |E_z|^2$. Solving (5) and (6) gives

dispersion relation in the short wavelength range relevant to the solar wind

$$\frac{\omega^2}{\Omega_e^2} = \bar{k}_z^2 \bar{k}_\perp^2 \frac{\left(\beta_\parallel + \beta_{\perp i}\right)}{\left(2 + \beta_{\perp i} + \beta_\perp\right)}, \tag{7}$$

with damping rate

$$\gamma_{KAW} = \frac{\sqrt{\pi}}{2} \omega \beta_\perp v_{t\parallel e}^2 \left.\frac{\partial f_{0e}}{\partial v_z}\right|_{v_z = \omega/k_z}. \tag{8a}$$

When $f_{0e}$ is a Maxwellian distribution, the damping rate is

$$\gamma_{KAW}^{Maxwellian} = -\frac{\sqrt{\pi}}{4} \omega \beta_\perp \zeta_e. \tag{8b}$$

The exponential factor, $\exp\left(-\zeta_e^2\right)$, was ignored in (8b) since from the KAW dispersion relation

(8), $\zeta_e \approx \bar{k}_\perp < 1$.

Cluster space correlation measurement analysis of the short wavelength range, near $k\rho_i \approx 2$, has measured small amplitude $\left(\delta B / B_0 < 10^{-2}\right)$ KAW [Sahraoui et al., 2010]. The short wavelength range, between the proton and electron gyroradius, is observed to have a scaling of the magnetic field power spectrum as $\delta B_{k\perp}^2 \sim k_\perp^{-2.8}$ [Alexandrova et al., 2009].

Recently it was found that quasilinear diffusion driven by KAW turbulence should establish a plateau-like distribution for the electrons [Rudakov et al., 2012]. The diffusion equation is

$$\frac{\partial f_{0e}}{\partial t} = \frac{\partial}{\partial v_z} D_e \frac{\partial f_{0e}}{\partial v_z}, \tag{9a}$$

with diffusion coefficient

$$D_e(v_z, v_\perp) = \pi \int \left| \frac{e}{m} E_{kz} + \frac{\mu}{m} i k_z B_{kz} \right|^2 \delta(\omega_k - k_z v_z) d\vec{k}, \tag{9b}$$

where the magnetic moment $\mu = T_{\perp e} / 2B_0$ is conserved. The diffusion coefficient for electrons was estimated from the measured KAW spectrum perpendicular magnetic field spectrum to be

$$D_e \sim \frac{2 \times 10^{-7}}{(v_z / v_{te})^4} \frac{MV_A^2}{m}.$$ It was estimated that the width of the plateau in the electron distribution

function is $v_{me} / v_{te} \sim (10^{-7} t)^{1/6} \sim 0.5$. The time of flight of the solar wind plasma to the earth is a few days. Figure 1 shows a numerical solution of the diffusion equation (9a) at time t=0 and $t = 10^5 s$. The initial distribution, which is assumed to be a Maxwellian, evolves to a distribution that has a plateau at velocities $v < v_{te}$. Since the minimum phase velocity of KAW is the Alfven velocity, for velocities $v < V_A$ the distribution remains Maxwellian. By making a numerical comparison of the derivative of the electron distribution, it is found that the damping rate is diminished by orders of magnitude. Thus the damping of KAW (8a) is drastically reduced. In the next section it will be similarly shown that the Landau damping of whistler waves is diminished as well.

### 3) Diminished Landau Damping of Whistlers in the Solar Wind

The general dispersion relation for whistlers, with frequency $\Omega_i << \omega < \Omega_e$, in a plasma with isotropic temperature $\left(T_\parallel \equiv T_\perp\right)$, is determined from the electron current $\vec{j}_k$ in a similar manner to KAW. From the x-component of Ampere's law, and under the restriction of long wavelength waves $\left(\bar{k}^2 << 1\right)$,

$$\bar{k}^2 \left(E_x + ik_x \phi\right) = -\frac{i\omega}{\Omega_e} E_y. \tag{10}$$

Equation (10) is equivalent to (1) in the limit that $\beta_{\perp i} \to \infty$. Physically this states that for whistlers, the ions provide a charge neutralizing background and are motionless.

Using the gauge condition for the field $E_x = -\left(E_z + ik_z\phi\right)\overline{k}_z / \overline{k}_\perp - ik_x\phi$ in (10), and subtracting from (3) gives the relation between the y and z component of the electric field

$$\frac{E_y}{E_z} = \frac{\alpha_Z}{\dfrac{i\omega}{\Omega_e}\dfrac{\overline{k}_\perp}{\overline{k}_z}}; \qquad \alpha_Z = \frac{2\omega^2}{\Omega_e^2\overline{k}_z^2}\frac{Z'}{2\beta_\parallel}\bigg/\left(1 - \frac{Z'}{2}\right). \qquad (11)$$

Notice that the parameter $\alpha_Z$ was defined in (11). Substituting $E_y$ from (11) in z-component of Ampere's Law (3) determines the electric field $E_z$

$$E_z = -ik_z\phi\frac{\overline{k}^2}{\overline{k}^2 + \alpha_Z}. \qquad (12)$$

Now substituting $E_y$ from (11) in x-component of Ampere's Law (10) solves for the potential in terms of $E_z$ and $E_x$, $ik_x\phi = \alpha_z\dfrac{\overline{k}_z}{\overline{k}_\perp\overline{k}^2}E_z - E_x$. And the potential can be used in the electric field $E_z$ (12) to determine the relation of $E_z$ and $E_x$

$$\frac{E_z}{E_x} = \frac{\overline{k}_z\overline{k}_\perp}{\overline{k}_\perp^2 + \alpha_Z}. \qquad (13)$$

The dispersion relation is found from the y-component of Ampere's Law (2), after substituting the relations (11) and (13)

$$\frac{\omega^2}{\Omega_e^2} = \frac{\alpha_Z\overline{k}^2\overline{k}_z^2}{\overline{k}_\perp^2 + \alpha_Z}. \qquad (14)$$

It is a remarkable feature that for both small and large arguments of the Z-function, $\zeta_e^2 = \omega^2\big/\overline{k}_z^2\Omega_e^2\beta$, the dispersion relation of whistlers is identical. This occurs because the parameter $\alpha_Z = \pm 1$ for large and small $\zeta_e$ respectively, which is seen by employing the asymptotic and power series expansions of the Z-function

$$\alpha_Z = \begin{cases} 1 + i\sqrt{\pi}\,\zeta_e \exp\left(-\zeta_e^2\right) & \zeta_e^2 >> 1 \\ -1 - \dfrac{i\sqrt{\pi}}{2\zeta_e} \exp\left(-\zeta_e^2\right) & \zeta_e^2 << 1 \end{cases}. \tag{15}$$

In these limits the dispersion relation (14) reduces to

$$\frac{\omega^2}{\Omega_e^2} = \begin{cases} \dfrac{\bar{k}^2 \bar{k}_z^2}{1 - \bar{k}_\perp^2} = \bar{k}_z^2 \bar{k}^2 & \zeta_e^2 = \bar{k}^2/\beta << 1; \quad \bar{k}_\perp^2 << 1 \\ \dfrac{\bar{k}^2 \bar{k}_z^2}{1 + \bar{k}_\perp^2} = \bar{k}_z^2 \bar{k}^2 & \zeta_e^2 = \bar{k}^2/\beta >> 1; \quad \bar{k}_\perp^2 << 1 \end{cases}. \tag{16}$$

It is clear from these two limits that $\zeta_e >> 1$ corresponds to small $\beta$, and $\zeta_e << 1$ corresponds to large $\beta$. Thus the general dispersion relation (14) is valid for the small and large $\beta$ limit as well. In the small $\beta$ (cold plasma) limit, the components of the electric field (11) and (13) are $E_{ky} = i\dfrac{\omega}{\Omega_e}\dfrac{E_{kx}}{\bar{k}^2}$, and $E_{kz} = E_{kx}\bar{k}_z\bar{k}_x$, respectively, as shown in Ganguli et al. [2010]. The resonant phase velocity, $\mathrm{v}_{res} \equiv \omega/k_z$, is small $\mathrm{v}_{res}/\mathrm{v}_{te} = \bar{k}/\beta^{1/2} < 1$, when $\beta$ is not too small as in the solar wind. As a consequence of the field relation (11), for high beta plasmas $\beta >> E_z/E_y = \bar{k}^2$ the parallel electric force small compared to $\mu\nabla\delta B_z$.

Since the KAW turbulence in the solar wind can create a distribution with a diminished derivative in the electron distribution function (see fig. 1), the damping rate of whistlers

$$\gamma_{whistler} = \sqrt{\pi}\,\omega\beta_\perp \frac{k_\perp^2}{k^2}\,\mathrm{v}_{t\|e}^2 \left.\frac{\partial f_0}{\partial \mathrm{v}_z}\right|_{\mathrm{v}_z = \omega/k_z}. \tag{17a}$$

should be diminished by the same factor as KAW. For a Maxwellian distribution function, the damping rate of whistlers is

$$\gamma_{Whistler}^{Maxwellian} = -\frac{\sqrt{\pi}}{2}\,\omega\beta_\perp \frac{k_\perp^2}{k^2}\,\zeta_e \tag{17b}$$

The exponential factor, $\exp\left(-\zeta_e^2\right)$, was ignored in (17b) since from the whistler dispersion relation for large beta $\zeta_e^2 \approx \overline{k}^2 / \beta << 1$.

The dispersion relation (14) and damping rate of whistlers (17) was obtained for an isothermal plasma, $\beta_\parallel = \beta_\perp$. More generally for $\beta_\parallel >> \overline{k}^2$, the whistler dispersion relation valid for an arbitrary non-Maxwellian distribution function in a plasma with anisotropic pressure (obtained by L. Rudakov from the drift-kinetic equation, to be published) is

$$
\frac{\omega^2}{\overline{k}^2 \overline{k}_z^2 \Omega_e^2} = \left(1 - \frac{k_\perp^2}{k^2}\frac{\beta_\perp^2 - \beta_\parallel^2}{2\beta_\parallel} - \frac{\overline{k}_z^2}{\overline{k}^2}\frac{\beta_\parallel - \beta_\perp}{2}\right)\left(1 - \frac{\beta_\parallel - \beta_\perp}{2}\right)
$$
$$
+ i\pi\frac{k_\perp^2}{k^2}\frac{\beta_\perp^2 + \beta_\parallel^2}{2\beta_\parallel}\left(1 - \frac{\beta_\parallel - \beta_\perp}{2}\right)\mathrm{v}_{t\parallel e}^2\left.\frac{\partial f_0}{\partial \mathrm{v}_z}\right|_{\mathrm{v}_z = \omega / k_z} .
$$

(18)

It is readily verified that for $\beta_\parallel = \beta_\perp$, the whistler dispersion (18) reduces to (16).

Whistlers may be generated from a variety of sources. They may be generated when the ion distribution is driven unstable by the KAW turbulence [Rudakov et al., 2012], or from Strahl electrons (Vinas et al., 2010). For cyclotron resonance of whistlers $\left(\omega << \Omega_e\right)$ with Strahl electrons, $\mathrm{v}_z = \Omega_e / k_z$,

$$
T_s = \frac{1}{2}m\mathrm{v}_z^2 = \frac{m}{2}\frac{\Omega_e^2}{k_z^2} = \frac{m}{2}\frac{c^2\Omega_e^2}{\omega_{pe}^2}\left(\frac{1}{\overline{k}_z^2}\right),
$$

(19)

where $T_s$ is the Strahl temperature. For typical solar wind parameters, $\frac{m}{2}\frac{c^2\Omega_e^2}{\omega_{pe}^2} = \frac{1}{2}MV_A^2 \sim 10\,eV$, and $T_s \approx 100 - 1000\,eV$, whistlers have a frequency $\omega / \Omega_e = \overline{k}_z^2 \sim 0.01 - 0.1$, with wavenumber $0.1 < \overline{k}_z < 0.3$.

However as whistlers propagate through the inhomogeneous solar wind, they acquire an ever-increasing perpendicular wavenumber through refraction. From the raytracing equation $dk_\perp/dt = -\partial \omega/\partial x$ (Stix, 1992), and the whistler dispersion $\omega = \bar{k}_\perp k_z M V_A / m \sim (B/n)$, the increase in $k_\perp$ can be estimated

$$\frac{\partial \omega}{\partial x} = \omega \frac{\partial}{\partial x} \ln\left(\frac{B}{n}\right) \sim \omega \frac{\Delta \ln(B/n)}{\Delta \tau V_{SW}}. \tag{20}$$

The gradient length, $\Delta x \sim V_{SW} \Delta \tau$, can be expressed in terms of time $\Delta \tau$ because the group velocity of whisters ($\sim 10^3 km/s$) is greater than the velocity of the solar wind ($V_{SW} \sim 500 km/s$), so the inhomogeneities of the solar wind can be considered stationary. Then eliminating the time variable from the ray-tracing equation, results in

$$\frac{\Delta k_\perp}{k_\parallel} \sim \Delta \ln(B/n) \frac{M V_A}{m V_{SW}} \bar{k}_\parallel \sim 0.2 \times 5 \times \bar{k}_\parallel \sim \bar{k}_\parallel. \tag{21}$$

In (21) it was estimated from satellite measurements of the magnetic field and density fluctuations that the inhomogeneity is just 20% [Vinas et al., 2010]. Since the whistler damping rate (17) increases with $k_\perp$, whistler waves would be damped without a sufficiently powerful source. However, with the modified electron distribution, whistlers are not damped so quickly and they can be longer lived in the solar wind.

## 4) Quasilinear Diffusion from Oblique Whistler Turbulence

Though measurements indicate that the short wavelength range of the solar wind turbulence spectrum consists of KAW, this does not preclude the existence of a spectrum of whistler turbulence. Indeed the results of the previous two sections demonstrate that whistlers should not be strongly damped in the solar wind. A possible path to whistler turbulence is through the three-wave interaction between whistlers and KAW. The three wave resonance condition $\omega_1^{whistler} = \omega_2^{whistler} + \omega_3^{KAW}$, and $\vec{k}_1^{whistler} = \vec{k}_2^{whistler} + \vec{k}_3^{KAW}$, can be readily satisfied for whistlers and kinetic Alfven waves. At oblique angles $\bar{k}_\perp^2 >> \bar{k}_\parallel^2$, whistler waves have a dispersion relation, $\omega^2 = \bar{k}_\parallel^2 \bar{k}_\perp^2 \Omega_e^2$, similar to KAW (7). With a measured KAW frequency of $\omega_{KAW} \approx 0.1\Omega_i$ and $\bar{k}_\parallel / \bar{k}_\perp \sim 0.1$, the 3-wave interaction changes the frequency of whistlers and parallel wavenumber only slightly, but $\bar{k}_{\perp 1}^{whistler} \pm \bar{k}_{\perp 3}^{KAW} \sim \bar{k}_{\perp 2}^{whistler}$. This scattering process will isotropize the whistler turbulence in the plane normal to the magnetic field.

Assuming that the whistler turbulence has a spectrum with a scaling of the magnetic field power spectrum similar to KAW, i.e. $\delta B_{k\perp}^2 \sim k_\perp^{-\nu}$, for $2 < \nu < 4$, then whistler turbulence would also result in the quasilinear evolution of the electron distribution function. From section 2 and 3, it is recalled that whistlers and KAW have resonance with the parallel electron distribution function when $v_z < v_{te}$ for sufficiently large $\beta$.

For large $\beta$ the parallel electric field (12) of whistlers is $E_{kz} = -ik_z \phi \bar{k}^{-2}$. It follows from (11) and (13) that at oblique angles $k_z / k_\perp < 1$, $\left| \delta B_x \right|^2 < \left| \delta B_y \right|^2$, $\left| \delta B_y \right|^2 = \left| \delta B_z \right|^2 = B_0^2 \left| e\phi / MV_A^2 \right|^2$, and thus the mirror force $-\mu_e \vec{\nabla} \delta B_z$ prevails over electric force as $\beta_\perp / \bar{k}^2$. Using these relations the diffusion coefficient resulting from the whistler turbulence is

$$
\begin{aligned}
D &= \pi \frac{v_{\perp e}^4}{4} \int k_z^2 \frac{\left| \delta B_z \left( k_z, k_\perp \right) \right|^2}{B_0^2} \delta(\omega_k - k_z v_z) dk_z k_\perp \, dk_\perp \\
&= \pi \frac{\beta_\perp^2}{4} \frac{\Omega_e^4 c^4}{\omega_{pe}^4} \int \frac{\left| \delta B_z \left( k_z, k_\perp \right) \right|^2}{B_0^2} k_z^2 \delta(\omega_k - k_z v_z) dk_z k_\perp \, dk_\perp \\
&= \pi \frac{\beta_\perp^2}{4} \frac{c^2 \Omega_e^2}{\omega_{pe}^2} \int N(\omega) \frac{\omega}{\Omega_e} d\omega \int \frac{\left| \delta B_{k\perp} \right|^2}{B_0^2 \bar{k}_\perp \left( c^2 / \omega_{pe}^2 \right)} \delta\left( \bar{k}_\perp - \frac{v_z}{v_0} \right) d\bar{k}_\perp \; .
\end{aligned}
\tag{22}
$$

The integration over $k_z$ was transformed to an integral over $\omega$ by using the dispersion relation,

$$k_z^2 = \omega^2 \omega_{pe}^4 / k_\perp^2 \Omega_e^2 c^4 = \omega^2 / \overline{k}_\perp^2 v_0^2 \quad k_z^2 = \omega^2 / \overline{k}_\perp^2 v_0^2 \quad \text{where} \quad v_0 = \sqrt{M/m}\, v_A.$$ The magnetic field

distribution is normalized such that $\int \left| \delta B_y \left( k_z, k_\perp \right) \right|^2 dk_z k_\perp \, dk_\perp = \dfrac{\omega_{pe}^2}{c^2} \int \left| \delta B_{k\perp} \right|^2 dk_\perp \int \dfrac{N(\omega)}{\Omega_e} d\omega$,

where $N(\omega)$ is the field distribution over $\omega$ and $\int \dfrac{N(\omega)}{\Omega_e} d\omega = 1$. The diffusion coefficient (22)

is similar to that of KAW (Rudakov et al., 2012), but the KAW coefficient has an extra factor of

$v_{te}^4 / v_{kaw}^4$ where $v_{kaw}^2 = (v_{te}^2 / 2 + T_{\perp i} / m)$.

If we assume a magnetic field spectrum of whistlers $\left| \delta B_{k\perp} \right|^2 = \left| \delta B \right|^2 \overline{k}_\perp^{-\nu} \left( c / \omega_{pe} \right)$, then

$$D = \pi \frac{\beta_\perp^2}{4} \left\langle \omega_{whistler} \right\rangle \frac{M V_A^2}{m} \frac{\left| \delta B_{whistler} \right|^2}{B_0^2} \left( \frac{v_0}{v_z} \right)^{\nu+1}. \qquad (23)$$

The average frequency $\left\langle \omega_{whistler} \right\rangle$ is determined from the integral over $\omega$ in equation (22). For an average whistler frequency $\left\langle \omega_{whistler} \right\rangle \geq 10^3 \left\langle \omega_{KAW} \right\rangle$, the magnetic field amplitude of whistlers need only be a fraction of KAW, $\left| \delta B_{whistler} \right|^2 \sim 10^{-3} \left| \delta B_{KAW} \right|^2$, to achieve a diffusion rate that is equivalent to KAW. With only this very small amplitude, whistler turbulence would also generate a plateau, as in Fig. 1 for KAW, within the time of flight of the solar wind plasma to the earth $\left( t \sim 10^5 \, s \right)$. For this reason, the possibility of even small amplitude of whistler turbulence should not be ignored. Of course larger amplitude turbulence of oblique whistlers is possible, but it would lead to increased quasilinear diffusion of low velocity electrons. This may occur under rare conditions, but the size of the plateau is still expected to be $v / v_{te} \sim 0.5$.

**5) Conclusion**

Previously it was shown that the KAW turbulence in the solar wind leads to the quasi-linear diffusion of the electron and ion distribution function and subsequently a diminished Landau damping rate of KAW. Additionally the established ion distribution can be unstable to whistler waves [Rudakov et al., 2012]. In this article, it is shown that in a sufficiently high $\beta$ plasma, whistlers have a parallel phase velocity that coincides with the plateau at $v < v_{te}$ created by the KAW turbulence, and thus will also have a diminished Landau damping rate. This enables the whistlers to survive since otherwise the propagation of whistlers through the inhomogeneous solar wind increases the perpendicular wavenumber resulting in their Landau damping. Though measurements have shown that KAW exist in the short wave length range of the solar wind turbulence spectrum, there is no credible evidence that whistlers do not exist. Many researchers believe that whistler turbulence should exist as well [Stawicki et al., 2001; Gary and Smith, 2009; Podesta et al., 2010]. Because of the diminished damping rate, whistler turbulence could also exist and with small amplitudes can lead to important consequences.

**Acknowledgements:**

This work is supported by Naval Research Laboratory base funds. One of the authors (L. Rudakov) acknowledges support from NSF grant AGS-1004270 at UCSD.

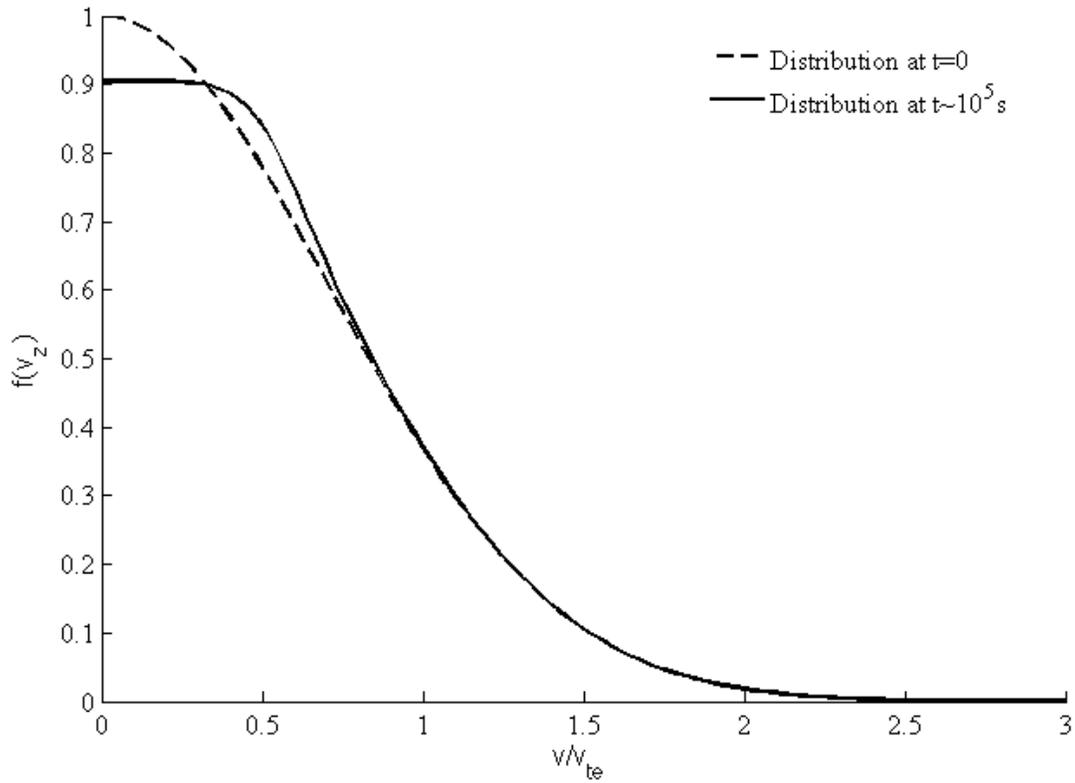

**Figure 1.** The initial Maxwellian distribution evolves through QL diffusion with KAW to a distribution that has a plateau at velocities $v < v_{te}$. For velocities $v < V_A$ the distribution remains Maxwellian, since the minimum phase velocity of KAW is $V_A$; this feature is not shown in the figure [Rudakov et al., 2011]. Thus the Landau damping of KAW (8) is diminished by a factor of 10 or more. The Landau damping of whistlers (17) is also diminished by the same factor.